\documentclass[fleqn,12pt]{wlscirep}

\pdfoutput=1
\usepackage{amssymb}
\usepackage{amsmath}
\usepackage{graphicx,booktabs,array}
\usepackage{url}
\usepackage{xcolor}
\usepackage{subcaption}
\usepackage{mwe}

\usepackage{setspace}
\title{Confirmation of the standard cosmological model from red massive galaxies $\sim600$ Myr after the Big Bang}

\author[1]{F.~Prada}
\author[2,3]{P.~Behroozi}
\author[4]{T.~Ishiyama}
\author[5]{A.~Klypin}
\author[1]{E.~P\'erez}

\affil[1]{Instituto de Astrof\'isica de Andaluc\'ia (IAA-CSIC), Granada, E-18008, Spain}
\affil[2]{Department of Astronomy and Steward Observatory, University of Arizona, Tucson, AZ 85721, USA}
\affil[3]{Division of Science, National Astronomical Observatory of Japan, 2-21-1 Osawa, Mitaka, Tokyo 181-8588, Japan}
\affil[4]{Digital Transformation Enhancement Council, Chiba University, 1-33, Yayoi-cho, Inage-ku, Chiba, 263-8522, Japan}
\affil[5]{Department of Astronomy, University of Virginia, Charlottesville, VA 22904, USA}

\begin{abstract}
\textbf{ARISING FROM} I. Labb\'e, P. van Dokkum, E. Nelson, R. Bezanson, K.A. Suess, J. Leja, G. Brammer, K. Whitaker, E. Mathews, M. Stefanon \& B. Wang. \textit{Nature} 616, 266 (2023)
\end{abstract}

\begin{document}

\flushbottom
\maketitle
\thispagestyle{empty}

\section*{MAIN TEXT}

In their recent study, Labb\'e et al. \cite{Labbe2023} used multi-band infrared images captured by the James Webb Space Telescope (JWST) to discover a population of red massive galaxies that formed approximately 600 million years after the Big Bang. The authors reported an extraordinarily large density of these galaxies, with stellar masses exceeding $10^{10}$ solar masses, which, if confirmed, challenges the standard cosmological model as suggested by recent studies\cite{Boylan-Kolchin2023, Lovell2023}. However, this conclusion is disputed. We contend that during the early epochs of the universe the stellar mass-to-light ratio could not have reached the values reported by Labb\'e et al. 
A model of galaxy formation based on standard cosmology provides support for this hypothesis, predicting the formation of massive galaxies with higher ultraviolet (UV) luminosity, which produce several hundred solar masses of stars per year and containing significant dust. These forecasts are consistent with the abundance of JWST/HST galaxies selected photometrically in the rest-frame UV wavelengths and with the properties of the recent spectroscopically-confirmed JWST/HST galaxies formed during that era. Discrepancies with Labb\'e et al. may arise from overestimation of the stellar masses, systematic uncertainties, absence of JWST/MIRI data\cite{Papovich2023}, heavy dust extinction affecting UV luminosities, or misidentification of faint red AGN galaxies at closer redshifts\cite{Labbe2023}. The current JWST/HST results, combined with a realistic galaxy formation model,
provide strong confirmation of the standard cosmology.

Fig. 1a shows the rest-frame UV luminosity function at $z=8$, 9, and 10, derived from galaxy samples selected photometrically in the rest-frame UV. Red symbols represent the latest measurements from the JWST\cite{Naidu2022, Donnan2023, Harikane2023}, while cyan symbols show those from the Hubble Space Telescope (HST)\cite{Oesch2018, Bouwens2021, Kauffmann2022}. We also show theoretical results based on well-tested and widely-used combination of cosmological $N-$body simulations with a galaxy formation algorithm that produces simulated galaxies \cite{Behroozi2019,Ishiyama2021}. Mock galaxy catalogues generated by this approach are called \textsc{Uchuu-UM}.
Solid lines indicate the luminosity functions obtained from our 
\textsc{Uchuu-UM} galaxy catalogues, while dashed lines are for the same galaxies with no internal dust extinction. The shaded region, only shown at $z=10$ as a reference, represents the uncertainty due to cosmic variance considering the observed JWST volume at these redshifts. The \textsc{Uchuu-UM} model accurately reproduces the observed JWST and HST UV luminosity functions over a range of five UV absolute magnitudes.

Fig. 1b shows the evolution of the cosmic UV luminosity and SFR density in galaxies integrated down to $M_{\rm UV}=-17$. Recent measurements with JWST are denoted by red symbols\cite{Donnan2023, Harikane2023}. The solid black line shows the results from  \textsc{Uchuu-UM} catalogues; the dotted line is the same without internal dust extinction. The shaded cyan region depicts the halo evolution model from Oesch et al.\cite{Oesch2018}, which predicts a rapid decline in density at $z > 8$. The extrapolated best-fit by Madau \& Dickinson\cite{MD2014} and the constant star-formation efficiency model from Harikane et al\cite{Harikane2023} are also shown as a reference. Our \textsc{Uchuu-UM} model is in good agreement with the evolution of UV/SFR density determined from JWST over the redshift range $z\sim7$$-$$14$.

The \textsc{Uchuu-UM} galaxy catalogues used in this analysis were generated by applying the \textsc{UniverseMachine}\cite{Behroozi2019} to assign galaxies to the dark matter halos in Shin-Uchuu and Uchuu1000-PL18 $N$-body simulations\cite{Ishiyama2021}. 
The \textsc{UniverseMachine} is a self-consistent empirical model that parameterises the galaxy SFR as a function of its host dark matter halo mass, mass accretion rate, and redshift. \textsc{UniverseMachine} was developed to accurately predict realistic global properties for individual high-redshift galaxies as observed by JWST\cite{Behroozi2020}, including UV luminosity, stellar mass, SFR, and dust extinction. Additionally, the model incorporates the evolution in the galaxy-halo relationship. 
Our galaxy catalogues cover redshifts from $z=0$ to 20. We use the same Planck cosmology for both simulations, with particle masses of $8.97 \times 10^5$ $h^{-1}$M$_{\odot}$ for Shin-Uchuu and $3.29 \times 10^8$ $h^{-1}$M$_{\odot}$ for UchuuPL18, and box sizes of 140 $h^{-1}$Mpc and 1000 $h^{-1}$Mpc, respectively.

We validate \textsc{Uchuu-UM} predictions by comparing them with nine galaxies detected with JWST\cite{Heintz2023,Fujimoto2023,Williams2023} and two galaxies with HST \cite{Bouwens2022}. These galaxies have measured spectra and span the redshift range of $7.7 < z < 9.5$. The weighted mean for the SFR - stellar mass relation of \textsc{Uchuu-UM} galaxies brighter than $M_{\rm UV}=-17$ at $z\sim8.5$ (individual galaxies are color-coded by its dust extinction A$_{\rm UV}$) agrees with the available sample of spectroscopically-confirmed JWST/HST galaxies (each symbol is colored according to the corresponding A$_{\rm UV}$ obtained from the galaxy spectra), see Fig. 1c. \textsc{Uchuu-UM} also provides a good explanation for the relationship between stellar mass and UV luminosity observed for those galaxies as shown in Fig. 1d. However, the eleven red massive galaxies in Labbé et al., denoted by blue cross symbols, exhibit anomalous properties, with an average stellar mass 50 times greater for their corresponding UV luminosities, resulting in a significantly higher mass-to-light ratio at this epoch. These galaxies show a significant deviation from the $3\sigma$ scatter of the \textsc{Uchuu-UM} galaxies, Fig. 1d. Labbé et al. considered the possibility of overestimated fiducial masses 
by factors of $>10-100$ due to faint red AGN. Recent findings\cite{Papovich2023} indicate that adding JWST-MIRI data reduces the derived stellar mass by 0.4 dex for most high-redshift galaxies. On the other hand, if we accept their stellar mass estimates, these galaxies likely suffer from significant dust extinction, requiring their intrinsic UV luminosities to be 15-40 times larger than observed, in order to reconcile with the stellar mass - UV luminosity relation shown in Fig. 1d. 

Fig. 2 depicts the cosmic evolution of stellar-mass density, with the black line representing the \textsc{Uchuu-UM} prediction for galaxies brighter than $M_{UV}=-17$ (equivalent to a few times $10^7$ solar masses) over the redshift range $5.5 < z < 10$. 
This prediction is in agreement with recent measurements compiled by Papovich et al.\cite{Papovich2023} (gray open symbols) 
and with the maximally allowable density assuming that galaxies experience a burst at $z=100$ followed by normal star formation (shaded region), as reported in their work. The green line indicates the empirical model of Finkelstein (2016), corrected by Papovich et al. when JWST/MIRI data constrain the stellar masses. Additionally, we compare Labbé et al.'s estimate of the stellar-mass density for galaxies above $10^{10}$ solar masses (blue cross, with systematic errors estimated from their Table 2) to the \textsc{Uchuu-UM} prediction for masses above $10^{10}$ solar masses (cyan dashed line). Notably, the Labb\'e et al. estimate is two orders of magnitude larger than the expected value at $z\sim8$. 


Labb\'e et al. results, if correct, pose a challenge to the standard cosmology. According to 
Boylan-Kolchin \cite{Boylan-Kolchin2023} they would imply near 100\% efficiency of converting ``normal" gas to stars. Another ``explanation" would be that Labb\'e et al. galaxies are extremely rare events \cite{Lovell2023}. None of those extreme assumptions are required if we assume recent JWST measurements as indicated in Figs. 1 and 2. Indeed, the efficiency of star formation in \textsc{Uchuu-UM} is $\sim 5-10\%$, typical for galaxies such as our Milky Way. There is no need to resort to extreme values statistics either: observed galaxies at high redshifts are typical objects. 

The strikingly large stellar-mass density observed in galaxies with stellar masses exceeding $10^{10}$ solar masses at $z\sim8$, as reported by Labb\'e et al., is a crucial point of conflict with the standard  galaxy formation model.
Our claims presented in this \textit{matters arising} research note add weight to the 
presence of potential systematic errors in Labb\'e et al.

\newcommand{\araa}{ARA\&A}   \newcommand{\aap}{Astron. Astrophys.}
\newcommand{\aj}{Astron. J.}         \newcommand{\apj}{Astrophys. J.}
\newcommand{\apjl}{Astrophys. J.}      \newcommand{\apjs}{Astrophys. J. Supp. Series}
\newcommand{\mnras}{Mon. Not. R. Astron. Soc.}   \newcommand{\nat}{Nature}
\newcommand{\sci}{Science}
\newcommand{\pasj}{Publ. Astron. Soc. Japan}     \newcommand{\pasp}{Publ. Astron. Soc. Pac.}
\newcommand{\procspie}{Proc.\ SPIE} \newcommand{\physrep}{Phys. Rep.}
\newcommand{\apss}{APSS}    \newcommand{\natast}{Nature Astronomy}
\newcommand{\solphys}{Sol. Phys.}
\newcommand{\actaa}{Acta Astronom}
\newcommand{\aaps}{Astron. Astrophys. Supp.}
\newcommand{\iaucirc}{IAU Circular}
\newcommand{\prd}{Phys. Rev. D}

\newpage

\section*{Data Availability}

The \textsc{Uchuu-UM} galaxy catalogues used in this work are publicly available at the \textsc{Skies \& Universes} site for cosmological simulations:
\url{https://www.skiesanduniverses.org/Simulations/Uchuu/}

\section*{Acknowledgements}
We acknowledge discussions with E. Alfaro, C-A. Dong-P\'aez, Y. Dubois, Y. Harikane, and M. Volonteri. We thank I. Labb\'e and Y. Harikane for their guidance on the calculation of UV absolute magnitudes. FP, AK, and EP acknowledge support from the Spanish MICINN funding grant PGC2018-101931-B-I00 and grant CEX2021-001131-S funded by MCIN/AEI/10.13039/501100011033.
PB was partially funded by a Packard Fellowship, Grant \#2019-69646. 
TI has been supported by IAAR Research Support Program in Chiba University Japan, 
MEXT/JSPS KAKENHI (Grant Number JP19KK0344 and JP21H01122), 
MEXT as ``Program for Promoting Researches on the Supercomputer Fugaku'' (JPMXP1020200109),
and JICFuS.
The Shin-Uchuu simulation was carried out on the Aterui II supercomputer at CfCA-NAOJ and Uchuu1000-PL18 was generated at the supercomputer Fugaku at the RIKEN Centerfor Computational Science (R-CCS). The construction of merger trees of those simulations were partially carried out on XC40 at the Yukawa Institute Computer Facility in Kyoto University.
We thank IAA-CSIC, CESGA, and RedIRIS in Spain for hosting the Uchuu data releases in the \textsc{Skies \& Universes} site for cosmological simulations. The \textsc{UniverseMachine} and \textsc{Uchuu-UM} data analysis have made use of the $skun6$@IAA-CSIC computer facility managed by IAA-CSIC in Spain (MICINN EU-Feder grant EQC2018-004366-P).

\section*{Author contributions}
 FP drafted the manuscript and contributed to the analysis efforts. AK and EP led the analysis of the \textsc{Uchuu-UM} galaxy catalogues, after PB applied the \textsc{UniverseMachine} to the Shin-Uchuu and Uchuu1000-PL18 $N$-body cosmological simulations and created the Uchuu-UM galaxy catalogues. TI ran the Shin-Uchuu and Uchuu1000-PL18 simulations and generated halo and merger trees for their dark matter halos. EP produced the figures. All authors contributed to the discussion and presentation of the results and reviewed the manuscript.

\section*{ Author information} The authors declare no competing financial interests. Correspondence and requests for materials should be addressed to FP at \texttt{f.prada@csic.es}

\newpage

\begin{figure*}
\centering
\includegraphics[scale=0.3]{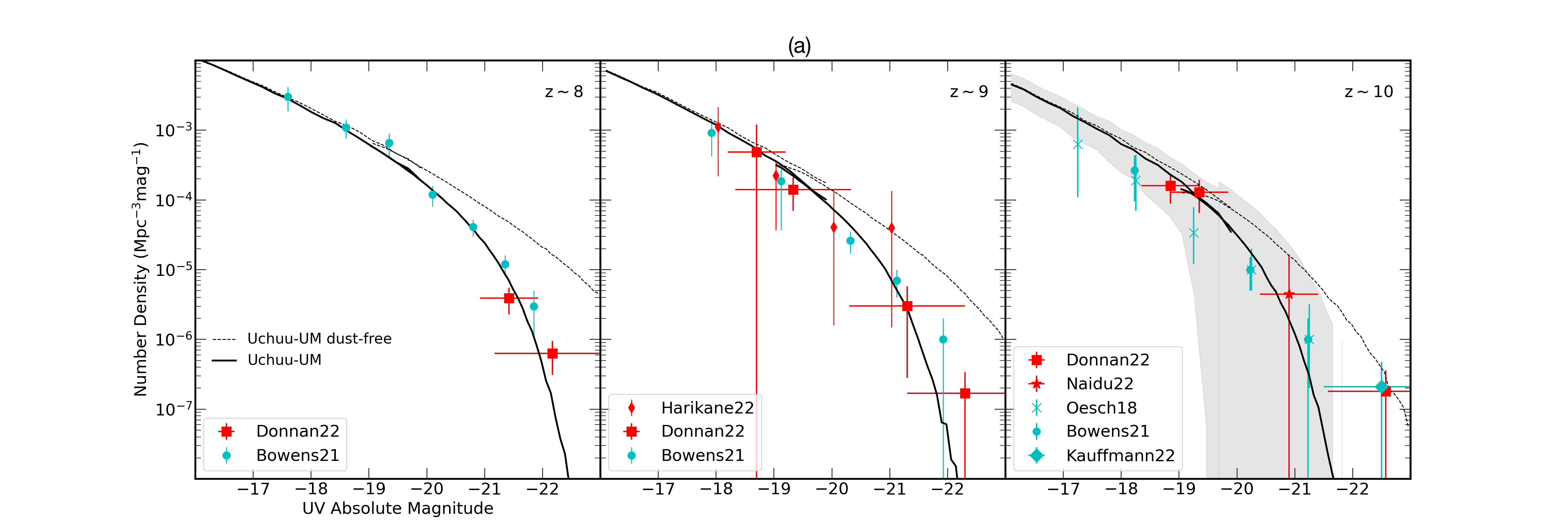}
\includegraphics[scale=0.26]{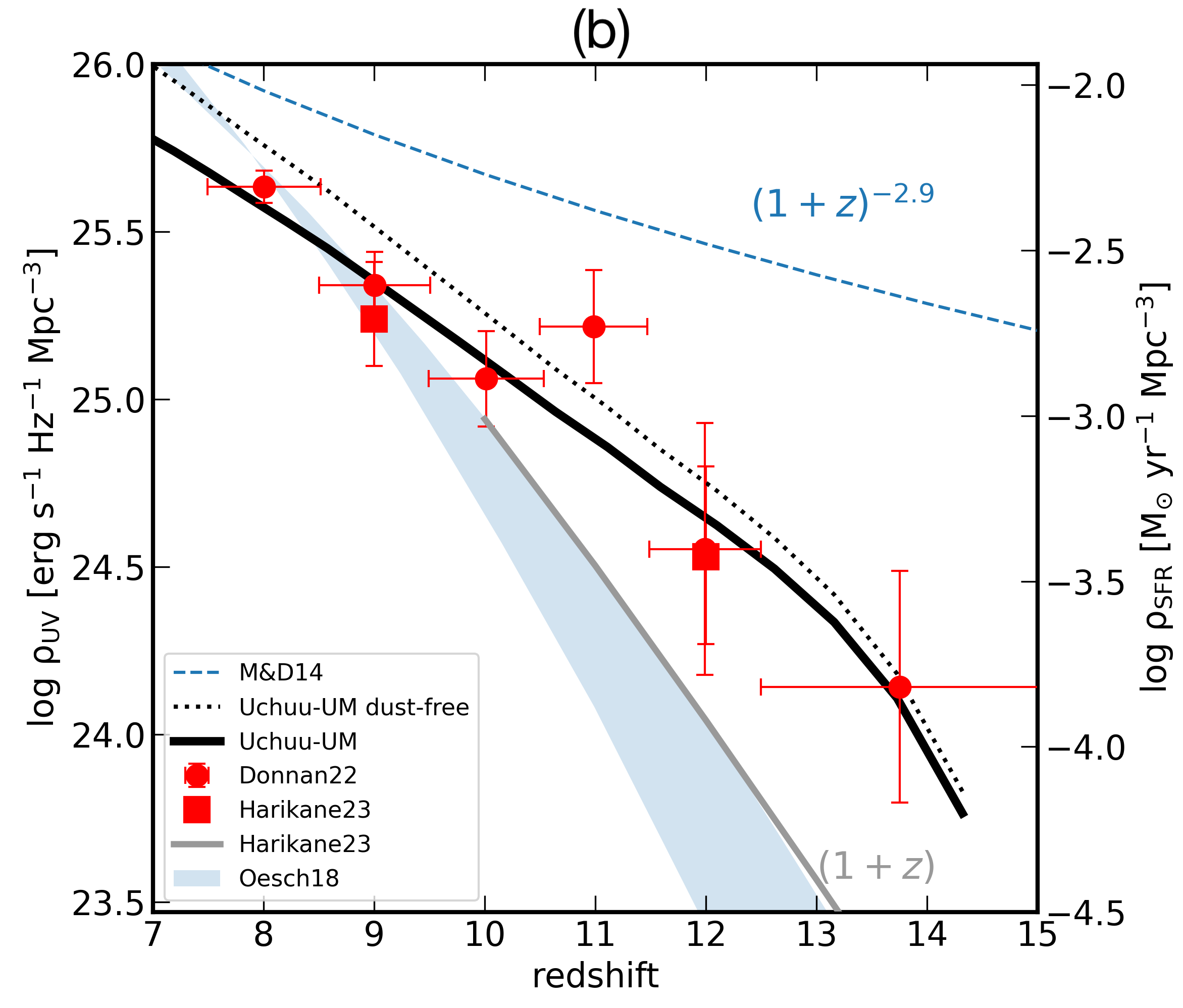}
\includegraphics[scale=0.31]{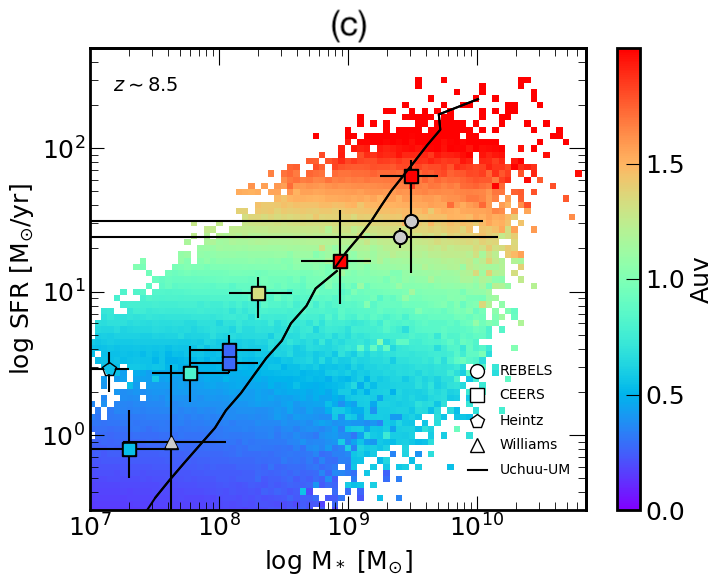}
\includegraphics[scale=0.31]{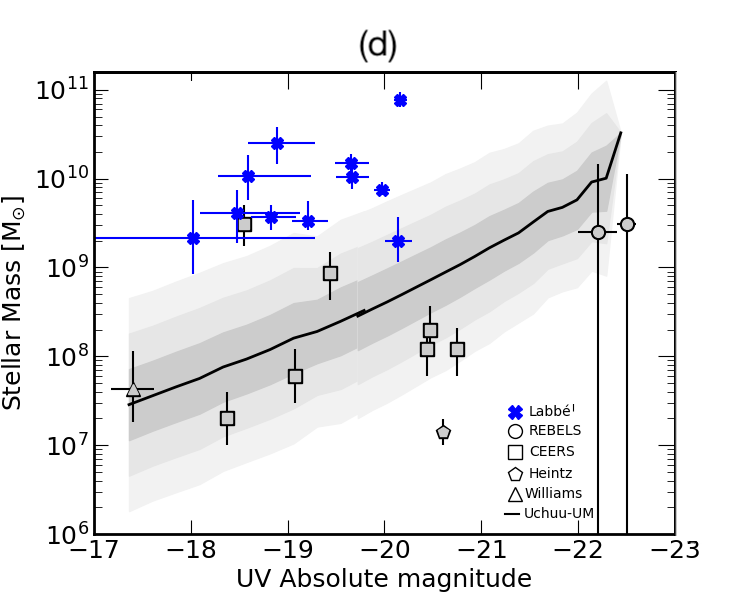}
\caption{
The \textsc{Uchuu-UM} galaxy formation model successfully explains the JWST\cite{Naidu2022, Donnan2023, Harikane2023} and HST\cite{Oesch2018, Bouwens2021, Kauffmann2022} UV luminosity function at redshifts 8, 9, and 10 (panel a), and predicts well the evolution of the observed UV/SFR density in JWST\cite{Donnan2023, Harikane2023} galaxies (panel b) and key global properties for spectroscopically-confirmed JWST/HST galaxies\cite{Heintz2023,Fujimoto2023,Williams2023,Bouwens2022} at a redshift of $z\sim8.5$ (panels c and d), including SFR vs. stellar mass (color-coded by extinction A$_{\rm UV}$) and stellar-mass vs. UV luminosity relations. The model addresses concerns raised by Labbé et al. \cite{Labbe2023} regarding red candidate galaxies at $z\sim8$ (indicated by blue crosses in panel d), which are 50 times more massive than expected based on their UV luminosity, and show a scatter above $3\sigma$ from the \textsc{Uchuu-UM} galaxies. This discrepancy could be due to overestimation of the stellar masses by a range of effects (systematic uncertainties, lack of JWST/MIRI data\cite{Papovich2023}, heavy dust extinction affecting the UV luminosities, or mistaken red AGN galaxies at closer redshifts\cite{Labbe2023}).
}
\label{fig:1}
\end{figure*}

\clearpage

\begin{figure}
\centering
\includegraphics[scale=0.5]{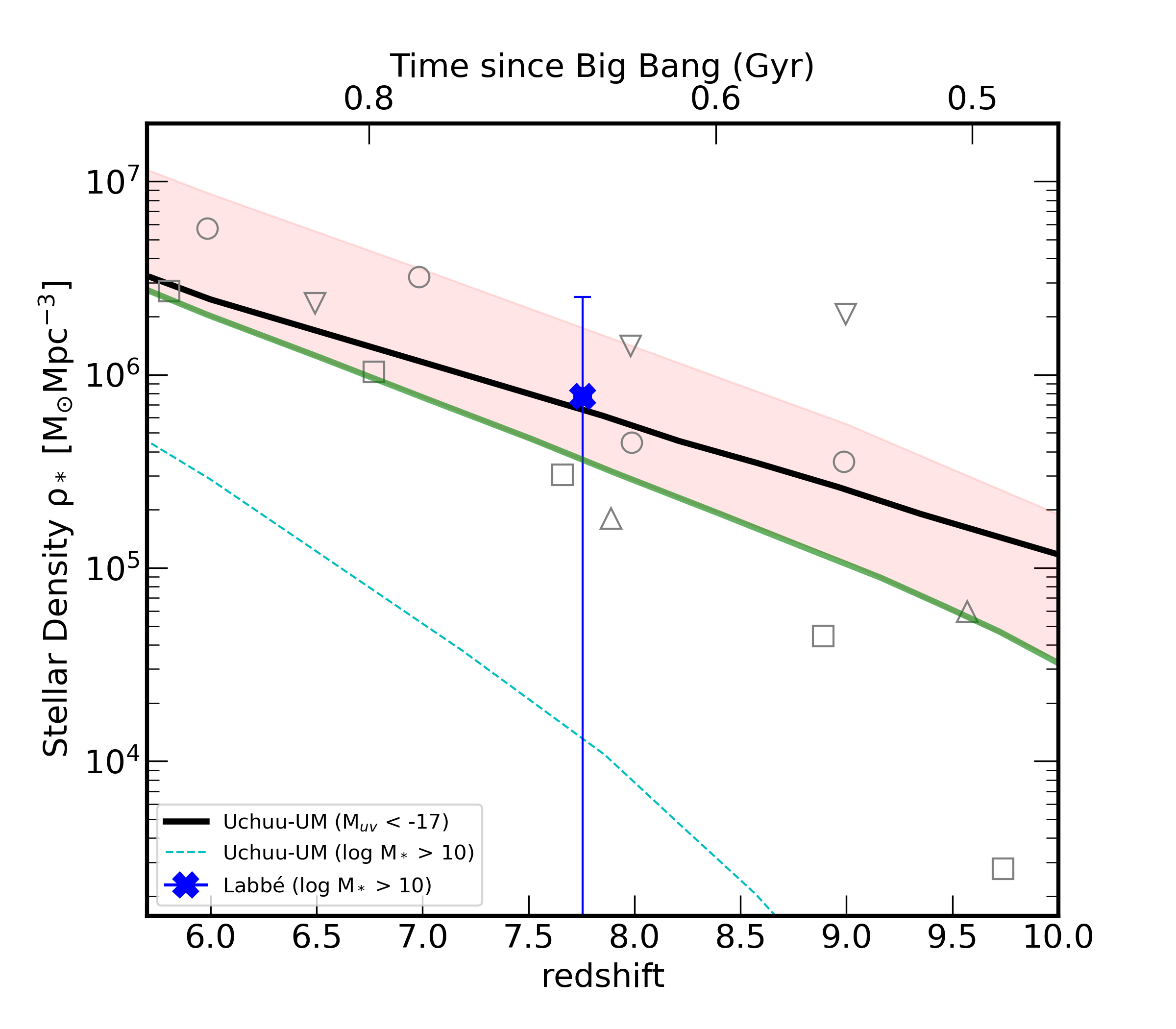}
\caption{
The black line depicts the \textsc{Uchuu-UM} stellar-mass density for galaxies more luminous than $M_{UV}=-17$ (equivalent to a few times $10^7$ solar masses).
This is in good agreement with recent measurements compiled by Papovich et al.\cite{Papovich2023} (gray open symbols) and with the maximally allowable density (reported in their work) assuming that galaxies experience a burst at $z=100$ followed by normal star-formation (shaded region). The green line indicates the empirical model of Finkelstein (2016) corrected by Papovich et al. when JWST/MIRI data have constrained the stellar masses. 
By contrast, Labbé et al.'s estimate of the stellar-mass density for galaxies above $10^{10}$ solar masses (blue cross, with systematic errors estimated from their Table 2) is compared to the \textsc{Uchuu-UM} prediction for masses above $10^{10}$ solar masses (cyan dashed line). Notice how Labb\'e et al. estimate is two orders of magnitude larger than the expected value at $z\sim8$. 
}
\label{fig:2}
\end{figure}

\clearpage




\end{document}